\documentclass[3p,times,procedia]{elsarticle}
\usepackage{nupha_ecrc}

\usepackage{mathtools}

\RequirePackage{lineno}
\volume{00}

\firstpage{1}

\journalname{Nuclear Physics A}

\runauth{P.~Braun-Munzinger, A.~Rustamov,  J.~Stachel}

\jid{nupha}

\jnltitlelogo{Nuclear Physics A}

\usepackage{amssymb}

\usepackage[figuresright]{rotating}

\begin{document}

\begin{frontmatter}

%% Instructions from Editor: Please use the following \dochead only in the preprint version (e-print arXiv etc.); 
%% use empty \dochead{} when submitting to Nuclear Physics A!
%\dochead{XXVIIth International Conference on Ultrarelativistic Nucleus-Nucleus Collisions\\ (Quark Matter 2018)}

\title{ Experimental results on fluctuations of conserved charges confronted with predictions from canonical thermodynamics}

\author[EMMI,HU]{P.~Braun-Munzinger }
\ead{p.braun-munzinger@gsi.de}
\author[GSI,NNRC,BSU,JINR]{A.~Rustamov}
\ead{a.rustamov@cern.ch}
\author[HU,EMMI]{J.~Stachel}
\ead{stachel@physi.uni-heidelberg.de}

\address[EMMI]{Extreme Matter Institute EMMI, GSI, Darmstadt, Germany}
\address[HU]{Physikalisches Institut, Universit\"{a}t Heidelberg, Heidelberg, Germany}
\address[GSI]{GSI Helmholtzzentrum f¸r Schwerionenforschung, Darmstadt, Germany}
\address[NNRC]{National Nuclear Research Center, Baku, Azerbaijan}
\address[BSU]{Baku State University, Baku, Azerbaijan}
\address[JINR]{Joint Institute for Nuclear Research, Dubna, Russia}

%\author[1,2,3,4]{Anar Rustamov}

%\address[1]{Azerbaijan National Academy of Sciences, Institute of Physics, Baku, Azerbaijan}
%\address[2]{Physikalisches Institut, Universit\"{a}t Heidelberg, Heidelberg, Germany}
%\address[3]{Extreme Matter Institute EMMI, GSI, Darmstadt, Germany}
%\address[4]{National Nuclear Research Center, Baku, Azerbaijan}
%\ead{a.rustamov@cern.ch}

\begin{abstract}
The study of multiplicity distributions of identified particles in terms of their higher moments is at the focus of contemporary experimental and theoretical studies. In a thermalized system,  combinations of these moments are directly related to the Equation of State  (EoS). The ultimate goal of the experimental measurements in relativistic nuclear collisions is, by systematic comparison to QCD and QCD inspired calculations, to probe the dynamics of genuine phase transitions between a hadron gas and the quark-gluon plasma.  However, the comparison between experiment and theory is far from trivial, because  several non-dynamical effects on fluctuations need to be controlled prior to a meaningful comparison to theoretical predictions. In this report we present quantitative estimates for these non-dynamical contributions using the Canonical Ensemble (CE) formulation of statistical mechanics. Together with analytical formulas we provide also results from Monte Carlo (MC) simulations within the CE and compare our predictions with the corresponding  measurements from the STAR experiment.
\end{abstract}


\end{frontmatter}

%\linenumbers

\section{Introduction}
\label{lintroduction}
One of the key goals of nuclear collision experiments at high energy is to  map the phase diagram of strongly interacting matter.  The most challenging part is the determination of the QCD
phase structure and the possible existence of a critical end point of a first order phase transition line, at which the matter undergoes a second-order phase transition. A promising tool to probe the presence of critical behavior is the study of fluctuations of conserved charges since, in a thermal system,  fluctuations are directly related to the EoS of the system under the study. One can  probe critical phenomena also at vanishing baryon chemical potential~\cite{LQCD1, Redlich1}. Moreover, the pseudo-critical temperature, reported from Lattice QCD (LQCD)~\cite{LQCD1}, is in agreement with the chemical freeze-out  temperature as extracted by comparing  Hadron Resonance Gas (HRG) model predictions~\cite{HRG} to the hadron multiplicities measured by ALICE. This agreement implies that strongly interacting matter, created in collisions of Pb nuclei at LHC energies, freezes out in close vicinity of the chiral phase transition line. Hence, singularities stemming from a second order phase transition can be captured also at vanishing net-baryon densities.
The current measurements,  by the STAR collaboration at RHIC, and  by ALICE at the LHC, have provided interesting and stimulating results. However, quantitative analysis of these measurements  is made difficult by the presence of non-critical effects such as volume or participant fluctuations and by correlations introduced by overall baryon number conservation. 

Conserved quantities fluctuate only in sub-regions of the available total phase space of the reaction. In statistical mechanics they are hence predicted within the Grand Canonical Ensemble (GCE)~\cite{StatLandau} formulation, where  only the average values of net-baryons are conserved~\cite{StatLandau}.  To compare theoretical calculations within GCE, such as HRG~\cite{HRG} and LQCD~\cite{LQCD1}, to experimental results, the requirements of GCE have to be achieved in experiments.  In experiments over the full acceptance, baryon number is conserved in each event, hence even in a limited acceptance its implications will be seen. Here, using the CE, we provide quantitative estimates of the implication of baryon conservation in a finite acceptance.
%The paper is organized in the following way: first we collect and summarize the notations and definitions used to compute cumulants. The next two sections deal with participant or volume fluctuations and the description of a simple model in which the effects of participant fluctuations can be quantitatively simulated. We apply these considerations in the following sections first to data from the ALICE experiment at the LHC, followed by applications to selected STAR data from the RHIC beam energy scan (BES). Next we discuss how to correct cumulant data for the effect of global conservation laws. In the final section we provide a conclusion and outlook.

\section{Fluctuations in GCE and CE}

In a thermal system with an ideal gas EoS, composed  of baryon/anti-baryon species with baryon numbers +1 and -1, GCE partition function  yields the uncorrelated Poisson distributions for baryons and anti-baryons, hence the net-baryon distribution has the following cumulants~\cite{ourModel}\footnote{The probability distribution of the difference of two random variables each generated from uncorrelated Poisson distributions is called Skellam distribution.}: 
\begin{equation}
\kappa_{n}(Skellam)=\left<n_{B}\right>+(-1)^n\left<n_{\bar{B}}\right>,
\label{netcumulants}
\end{equation}
where  $\left<n_{B}\right>$ and $\left<n_{\bar{B}}\right>$ denote the first cumulants (mean numbers) of baryons and anti-baryons, respectively. 
Eq.~(\ref{netcumulants})  implies that ratios of even-to-even and odd-to-odd cumulants of net-baryons are always unity,
while the ratios of odd-to-even cumulants depend on mean multiplicities.
% or alternatively on $\mu/T$:
\begin{alignat}{2}
%&\frac{\kappa_{2n}}{\kappa_{2k}} =\frac{\left<N_{B}\right>+\left<N_{\bar{B}}\right>}{\left<N_{B}\right>+\left<N_{\bar{B}}\right>} = 1 \nonumber\\
%&\frac{\kappa_{2n+1}}{\kappa_{2k+1}} = \frac{\left<N_{B}\right>-\left<N_{\bar{B}}\right>}{\left<N_{B}\right>-\left<N_{\bar{B}}\right>}= 1 \\
&\frac{\kappa_{2n+1}}{\kappa_{2k}}=\frac{\left<n_{B}\right>-\left<n_{\bar{B}}\right>}{\left<n_{B}\right>+\left<n_{\bar{B}}\right>}.
%= tanh\left(\frac{\mu}{T}\right) 
%\frac{\kappa_{2n+1}}{\kappa_{2k}}=\frac{<N_{B}>-<N_{\bar{B}}>}{<N_{B}>+<N_{\bar{B}}} = tanh(\frac{\mu}{T})
\label{conserv2}
\end{alignat}

Hitherto, the above conditions are used as baseline for net-baryon fluctuations. However, this can lead to misleading conclusions because, apart from dynamical fluctuations induced by critical phenomena, deviations from this baseline may be driven by non-dynamical contributions. Recently we demonstrated that fluctuations of participating nucleons from event-to-event significantly  distort measured event-by-event fluctuation signals~\cite{ourModel}.
At low energies\footnote{We note that at LHC energies, where mean numbers of net-baryons measured at mid-rapidity are zero, contributions from participant fluctuations to second and third cumulants of net-baryon distributions are vanishing.}, participant fluctuations always increase the measured dynamical fluctuations up to the third cumulant of net-proton distributions. In contrast, starting from the fourth cumulant, they can in fact decrease the signal.
\begin{figure}[htb]
\centering
 \includegraphics[width=0.45\linewidth,clip=true]{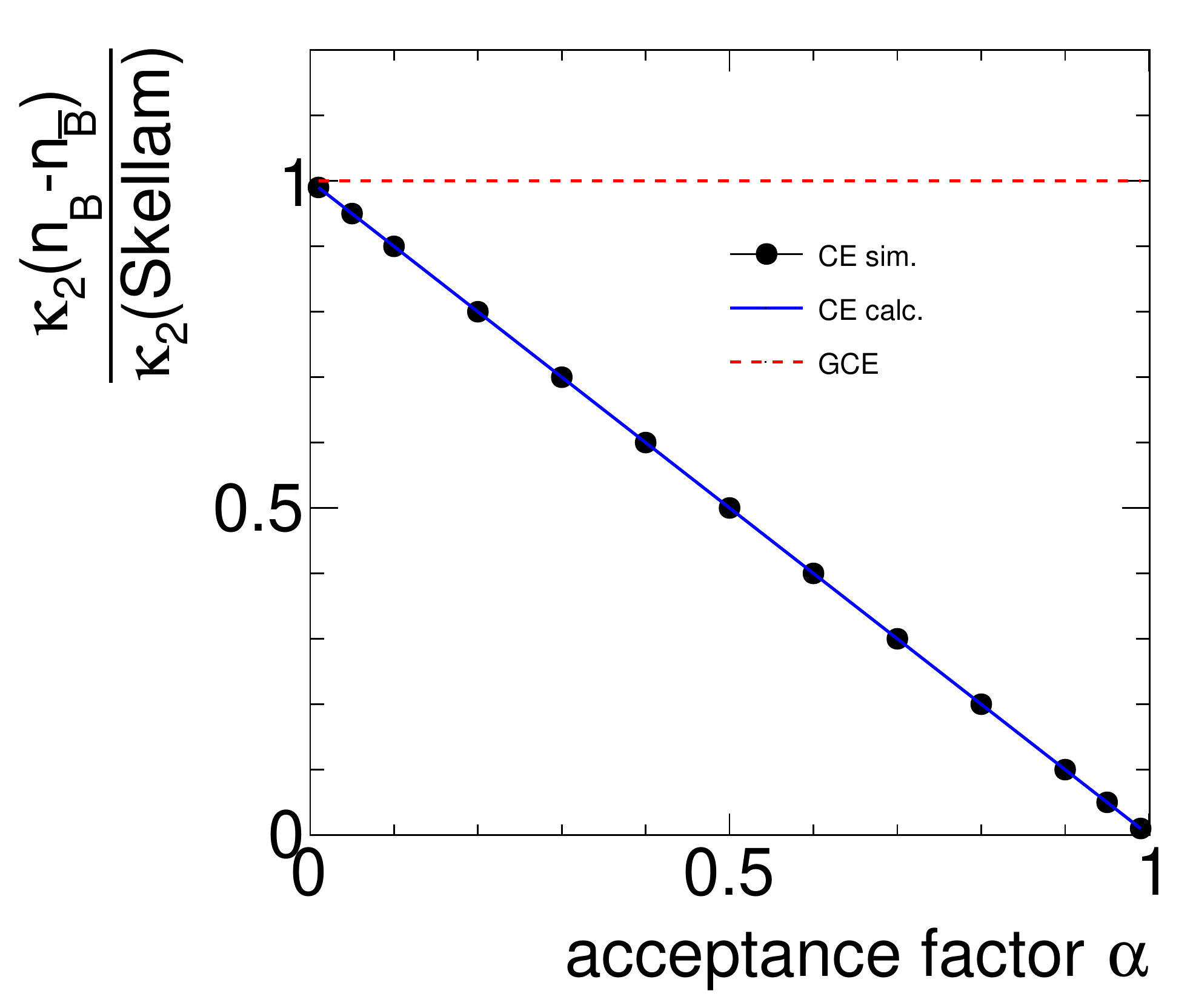}
 \includegraphics[width=0.45\linewidth,clip=true]{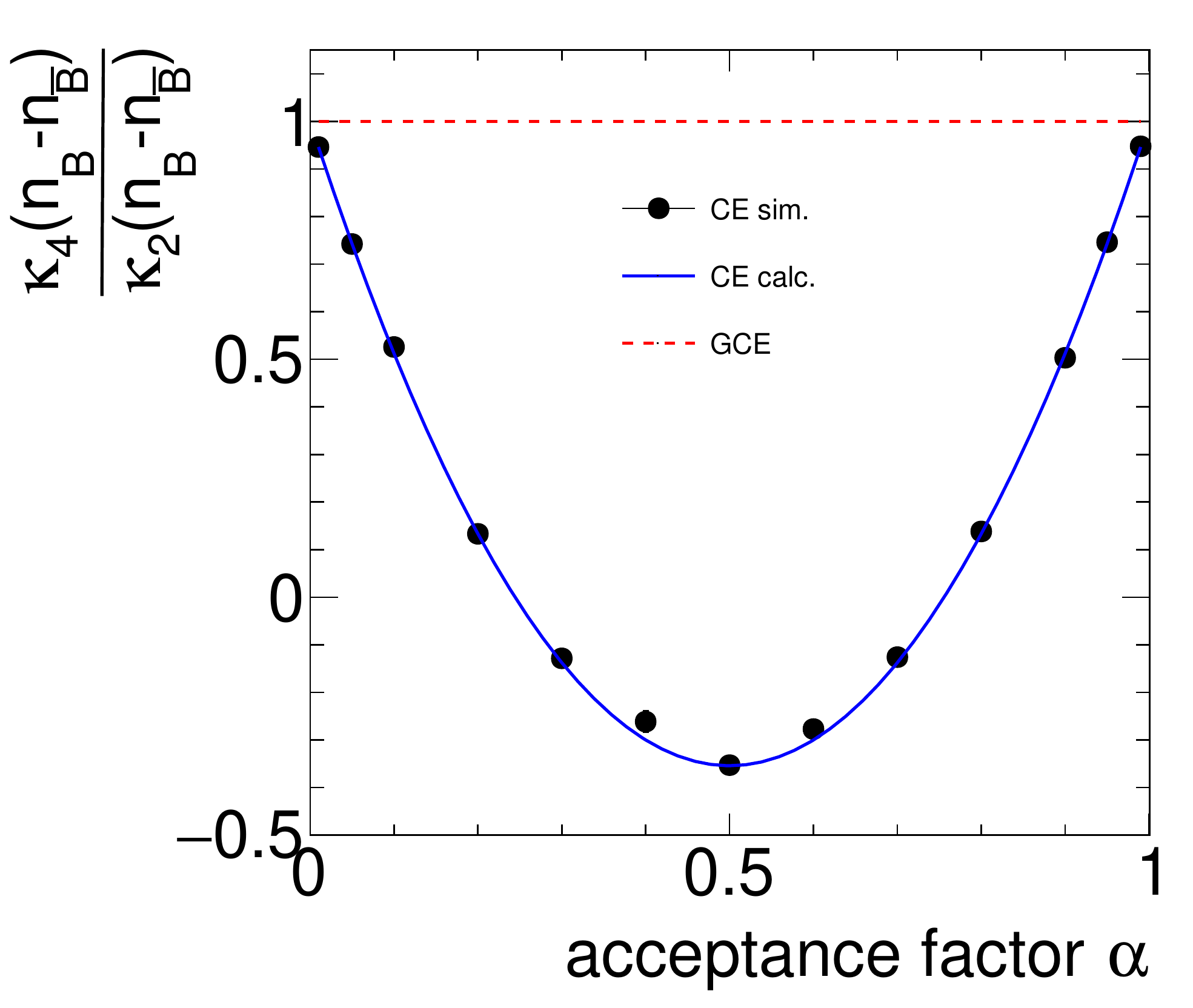} 
 \caption{Normalized cumulants of net-baryons from $5\times 10^{8}$ generated MC events in CE are presented with the black circles. The blue lines are calculated with Eqs.~(\ref{cumulantsCE1}, \ref{cumulantsCE3}), while the red dashed lines represent the GCE baseline.}
\label{fCE}
\end{figure} 
Below, we consider the CE partition function to investigate effects of exact baryon number conservation. It is
\begin{equation}
Z_{CE}(V,T,B)=\sum_{N_{B}=0}^{\infty}\sum_{N_{\bar{B}}=0}^{\infty}\frac{(\lambda_{B}z_{B})^{N_{B}}}{N_{B}!}\frac{(\lambda_{\bar{B}}z_{\bar{B}})^{N_{\bar{B}}}}{N_{\bar{B}}!}\delta({N_{B}-N_{\bar{B}}-B})= \left(\frac{z_{B}}{z_{\bar{B}}}\right)^{\frac{B}{2}}I_{B}(2\sqrt{z_{B}z_{\bar{B}}})\bigg\rvert_{\lambda_{B,\bar{B}}=1},
\label{zCE}
\end{equation}
where $I_{B}$ denotes the modified Bessel function, $\lambda_{B,\bar{B}}$ are fugacities and $z_{B,\bar{B}}$ stand for single particle partition functions of baryons and anti-baryons respectively.
%$N_{B,\bar{B}}$ and $z_{B,\bar{B}}$ stand for GCE multiplicities and single particle partition functions of baryons and anti-baryons.
The $\delta$ function  in Eq.~(\ref{zCE}) guarantees that, in each event, the net number of baryons is fixed, i.e, net-baryons do not fluctuate from event-to-event.  In order to get finite fluctuations for net-baryons, distributions of baryons and anti-baryons have to be folded with the experimental acceptance. Following the  acceptance folding strategy developed in~\cite{ourModel, koch-conserv}, we get the following cumulants for net-baryons in CE:
%\begin{alignat}{2}
%&\left[\frac{\kappa_{2}(n_{b}-n_{\bar{b}})}{\left<n_{b}+n_{\bar{b}}\right>}\right]_{CE}=\left[\frac{\kappa_{2}(n_{b}-n_{\bar{b}})}{\left<n_{b}+n_{\bar{b}}\right>}\right]_{GCE}-\alpha \label{cumulantsCE1}, \\
%&\left[\frac{\kappa_{3}(n_{b}-n_{\bar{b}})}{\kappa_{2}(n_{b}-n_{\bar{b}})}\right]_{CE}=\left[\frac{\kappa_{3}(n_{b}-n_{\bar{b}})}{\kappa_{2}(n_{b}-n_{\bar{b}})}\right]_{GCE}\times\left(1-2\alpha\right) \label{cumulantsCE2},\\
%&\left[\frac{\kappa_{4}(n_{b}-n_{\bar{b}})}{\kappa_{2}(n_{b}-n_{\bar{b}})}\right]_{CE}=\left[\frac{\kappa_{4}(n_{b}-n_{\bar{b}})}{\kappa_{2}(n_{b}-n_{\bar{b}})}\right]_{GCE}-6\alpha(1-\alpha)\times F \label{cumulantsCE3},
%\end{alignat}
\begin{alignat}{2}
&\left[\frac{\kappa_{2}(n_{B}-n_{\bar{B}})}{\kappa_{2}(Skellam)}\right]_{CE}=1-\alpha \label{cumulantsCE1}, \\
&\left[\frac{\kappa_{3}(n_{B}-n_{\bar{B}})}{\kappa_{2}(n_{B}-n_{\bar{B}})}\right]_{CE}=\left[\frac{\left<n_{B} - n_{\bar{B}}\right>}{\left<n_{B} + n_{\bar{B}}\right>}\right]_{CE}\times\left(1-2\alpha\right) \label{cumulantsCE2},\\
&\left[\frac{\kappa_{4}(n_{B}-n_{\bar{B}})}{\kappa_{2}(n_{B}-n_{\bar{B}})}\right]_{CE}=1-6\alpha(1-\alpha)\times F \label{cumulantsCE3},
\end{alignat}
where $\left<n_{B}\right>$ and $\left<n_{\bar{B}}\right>$ are the mean numbers of baryons and anti-baryons inside the acceptance and  $F$ is defined as:
\begin{equation}
F=1-2\frac{\left<N_{B}\right>_{CE}\left<N_{\bar{B}}\right>_{CE}}{\left<N_{B}+N_{\bar{B}}\right>_{CE}}\left(\frac{\left<N_{B}\right>_{GCE}\left<N_{\bar{B}}\right>_{GCE}}{\left<N_{B}\right>_{CE}\left<N_{\bar{B}}\right>_{CE}}-1\right),
\label{conserv2}
\end{equation}
with $N_{B,\bar{B}}$ referring to  baryons and anti-baryons in the full phase space.
We further note that the $\alpha$ parameter in Eqs.~(\ref{cumulantsCE1}-\ref{cumulantsCE3})  refers to the fraction of baryons falling into the experimental acceptance. 
We first generate the number of baryons and anti-baryons  from the  probability distributions encoded in the CE partition function (cf. Eq.~\ref{zCE}).  
%As mentioned above, although we get different numbers of baryons and anti-baryons in each event, their difference is always the same and equals $B$. 
Next, we randomly select the number of baryons and anti-baryons either with the binomial distribution or using  rapidity and transverse momentum spectra of baryons and anti-baryons. The results for the normalized values of $\kappa_{2}$ and $\kappa_{4}$ of net-baryons, as a function of accepted fraction of baryons, are presented in Fig.~\ref{fCE}, where the lines are analytical calculations with Eqs~(\ref{cumulantsCE1}, \ref{cumulantsCE3}).\footnote{In this simulation we used $\langle N_{B}\rangle = 370$ and $\langle N_{\bar{B}}\rangle = 20$ for baryons and anti-baryons respectively.} 

Although with a different starting point, very similar results were obtained earlier in~\cite{koch-conserv}. The effects of baryon number conservation were also considered in~\cite{Schuster:2009jv,Keranen:2004eu}.

\section{Confronting experimental results}
\begin{figure}[htb]
\centering
 \includegraphics[width=0.45\linewidth,clip=true]{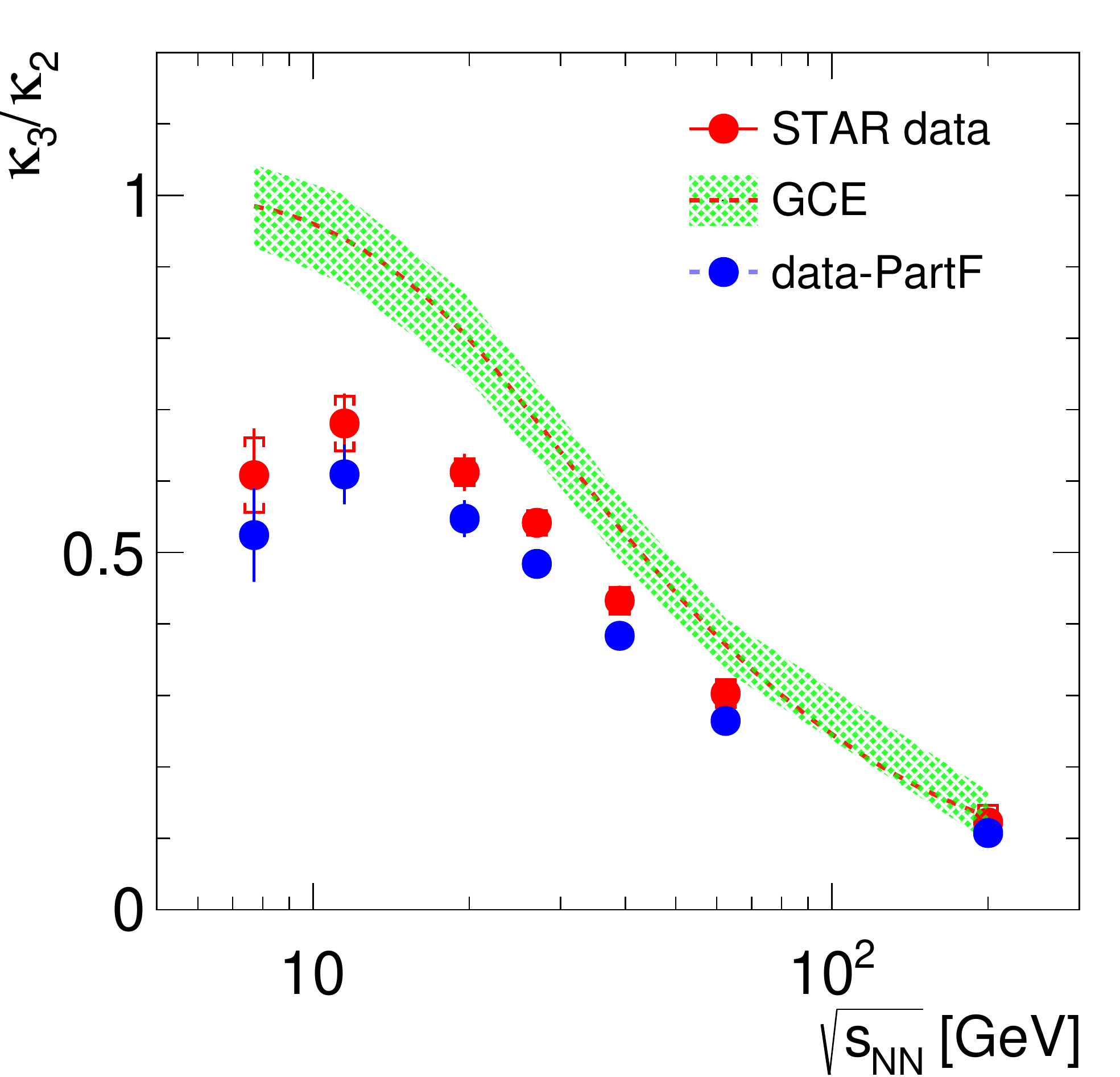}
 \includegraphics[width=0.45\linewidth,clip=true]{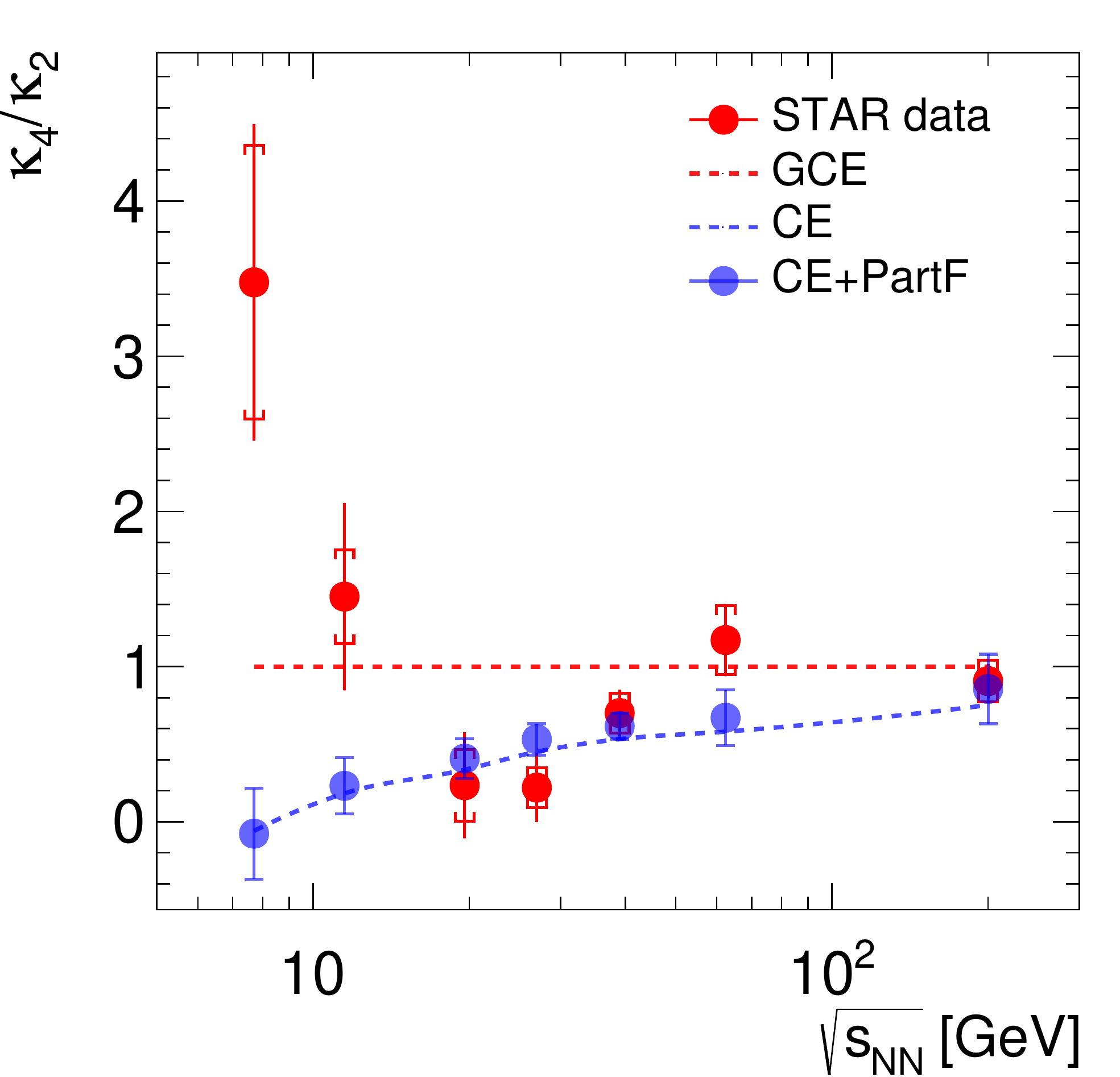} 
 \caption{Left panel: $\kappa_{3}/\kappa_{2}$ measurements from STAR (the red circles) and their corrected values for participant fluctuations (the blue symbols). Right panel: $\kappa_{4}/\kappa_{2}$ measurements from STAR (the red circles) compared to our predictions (the blue symbols). The blue dashed line corresponds to our predictions without participant fluctuations. The red dashed lines represent the GCE baseline. }
\label{figSTAR1}
\end{figure}

Next we present predictions for cumulants of net-protons at the RHIC BES energies. For this purpose, by using the energy dependence of $\kappa_{3}/\kappa_{2}$, as measured for net-protons by STAR~\cite{STARDATA}, we first fix the $\alpha$ parameter entering Eqs.~(\ref{cumulantsCE1} - \ref{cumulantsCE3}). As seen from the left panel of Fig.~\ref{figSTAR1} (red circles) the deviation of experimental measurements form the GCE line increases with decreasing energy. Moreover, the experimental measurements are always below the GCE values. This means that the conservation laws, which decrease the amount of fluctuations  (cf. Eqs.~\ref{cumulantsCE1} - \ref{cumulantsCE3}) are much stronger than effects due to fluctuations of participating nucleons.  Participant fluctuations  push  the  $\kappa_{3}/\kappa_{2}$ data in the opposite direction. Using the the procedure reported in~\cite{ourModel} we present  STAR data corrected for possible fluctuations of participant nucleons (blue circles in Fig.~\ref{figSTAR1}). Next, inserting the numerical values of the corrected  $\kappa_{3}/\kappa_{2}$ data into Eq.~(\ref{cumulantsCE2}), we obtain the energy dependence of the $\alpha$ parameter. Finally, using these values of $\alpha$ we present in the right panel of Fig.~\ref{figSTAR1}, with the blue dashed line, excitation function of $\kappa_{4}/\kappa_{2}$  as calculated using Eq.~\ref{cumulantsCE3}.  We further add contributions from participant fluctuations , which are presented by light blue circles.  As seen from Fig.~\ref{figSTAR1}, besides the points at $\sqrt{s_{NN}}$ = 7.7A and 11.5A GeV our predictions quantitatively reproduce the trend of $\kappa_{4}/\kappa_{2}$. Similar conclusions we get for the energy dependence of $\kappa_{1}/\kappa_{2}$ and $\kappa_{1}/\kappa_{3}$ (not presented here).  We hence conclude that, above $\sqrt{s_{NN}}$=11.5A GeV,  the experimentally observed deviations from the GCE baselines  can be  described by the combined effects of participant fluctuations and global conservation laws, the latter being dominant. Finally, we remark that the measurements  from the ALICE experiment can also be explained by the baryon number conservation~\cite{RustamovQM17}. 
   
\section{Conclusions}
We studied the effects of global conservation laws on fluctuations of net-baryon number. Together with analytic formulas we developed MC methods to simulate events in the CE. Above 11.5 GeV, the deviations from the Skellam distribution reported by STAR are consistently described with baryon number conservation and fluctuations of participating nucleons. A dramatic exception are the STAR results on $\kappa_{4}/\kappa_{2}$ below $\sqrt{s_{NN}}= 11.5$ GeV. The measured second cumulants of net protons at ALICE can also be accounted for quantitatively by conservation laws. Our results  will be relevant for the research programs at facilities such as FAIR at GSI and NICA at JINR. Near future challenges will be precision measurements of higher moments at RHIC and LHC and their connection to fundamental QCD predictions.
\section*{Acknowledgments}
This work is part of and supported by the DFG Collaborative Research
Centre "SFB 1225 (ISOQUANT)".

\end{document}